\documentclass{article}

\usepackage{graphicx}
\usepackage{color}
\usepackage{amsmath,amssymb}
\usepackage{setspace}

\begin{document}

\begin{center}
{\large Magnetic properties of AlB$_2$-type holmium silicides and germanides} \\ \vspace{10pt}

{\small G.\,Eguchi$^1$, R.\,Matsumoto$^1$, K.\,Terashima$^1$, and Y.\,Takano$^{1,\,2}$ \\

$^1$National Institute for Materials Science, 1-2-1 Sengen, Tsukuba, Ibaraki 305-0047, Japan \\
$^2$University of Tsukuba, 1-1-1 Tennodai, Tsukuba, Ibaraki 305-8577, Japan \\

\noindent Email: eguchi.gaku@nims.go.jp \\
(\noindent Date: \today) \\
}
\end{center}


\noindent \textbf{Abstract} \\
Discovery of the large magnetocaloric effect in HoB$_2$ has highlighted the practical advantage of heavy rare-earth ions. Other holmium compounds are of interest, and we here report the synthesis and the magnetic properties of HoSi$_{1.67}$ and HoGe$_{1.67}$ which form the same AlB$_2$-type structure but with vacancies. They are found to show the antiferromagnetic order with the N\'eel temperature 17.6(2)\,K for HoSi$_{1.67}$ and 9.9(2)\,K for HoGe$_{1.67}$, and the magnetic entropy changes at the temperature are 0.05(1)\,J/cm$^3$K for HoSi$_{1.67}$ and 0.08(1)\,J/cm$^3$K for HoGe$_{1.67}$. Magnetic orders were suppressed by replacing vacancies with nickel, resulting in an increase of magnetic entropy changes. Distance between the in-plane Ho$^{3+}$ ions appears to be an important parameter leading to the transition between the antiferromagnetic (HoSi$_{1.67}$) and the ferromagnetic (HoB$_2$) order. The finding may aid the exploration of other heavy rare-earth compounds for similar applications.

\clearpage
\noindent  \textbf{Introduction} \\
Efficient refrigeration is one of the keys for reducing energy consumption and magnetocooling has the potential to make a significant impact on a global scale. Magnetic materials are used as coolants, and those with large magnetic moments are advantageous to increase the cooling capacity. For this purpose heavy rare-earths are particularly suitable as they are known to have the large local moments.

\begin{figure}[ht]
\centering
\includegraphics[width=\textwidth]{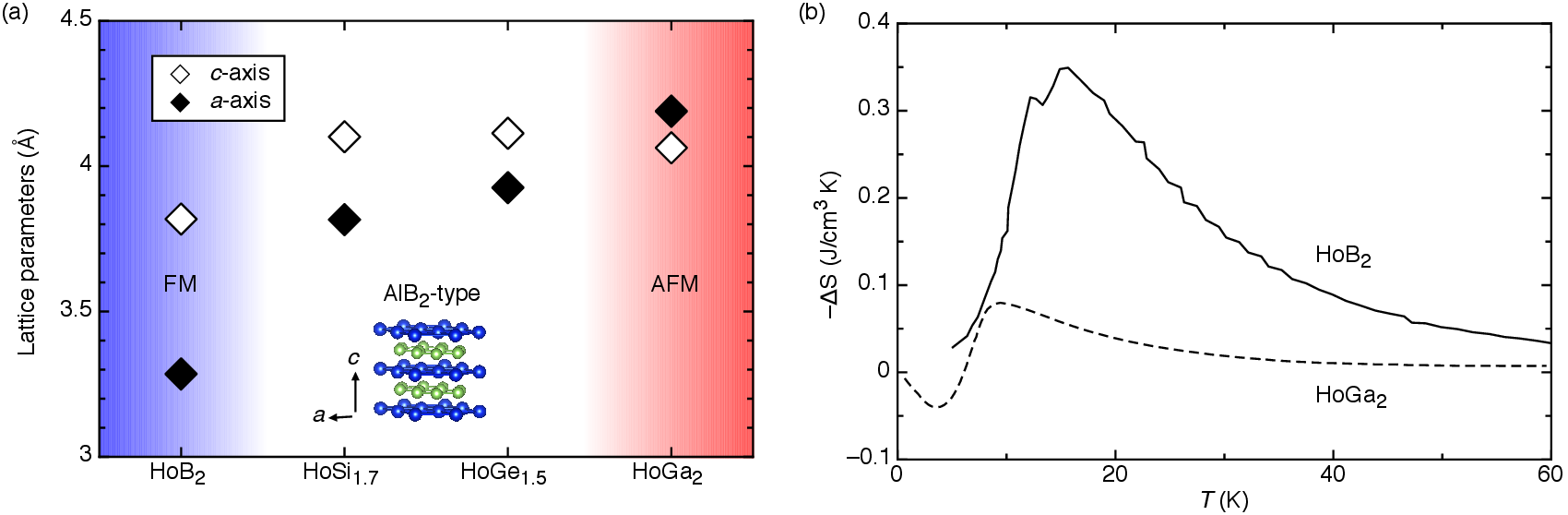}
\caption{\textbf{The AlB$_2$-type binary holmium compounds found in the literature.} (a)\,Lattice parameter and magnetic order. HoB$_2$ is known to show the ferromagnetic order (FM) and HoGa$_2$ the antiferromegnatic order (AFM). Inset is a schematic of the AlB$_2$-type structure\,[27]. (b) Magnetic entropy changes from 5\,T to 0\,T in the HoB$_2$ and in the HoGa$_2$\,[22, 23].}\label{fig1}
\end{figure}

Hydrogen liquefaction is one of the applications\,[1, 2]. Recently, large magnetocaloric effect was found in the HoB$_2$ near the boiling point of hydrogen\,[3]. It was proposed with the help of machine learning, and the finding highlights surprisingly rich functionalities that are still hidden in a simple rare-earth compound. Apparently the local moment on the Ho$^{3+}$ ion, which is known to have the large value by a factor of 10.5 compared to the Bohr magneton ($\mu_{\rm{B}}$)\,[4], harbor the changes. Such large local moments are common to holmium compounds, and therefore, we reviewed the magnetic properties of binary Ho$X_2$. One of the major categories is the Laves phase ($X=$Al, Mn, Fe, Co, Ni, Ru, Rh, Re, Os, Ir, Pt) and they are known to show the ferromagnetic order\,[5-9]. Others ($X=$H, C, Si, Cu, Zn, Ga, Ge, Ag, Au) take various crystal structures and they are known to show the antiferromagnetic order\,[10-18]. The above situation highlights peculiarity of the ferromagnetic order in the HoB$_2$. It forms so-called AlB$_2$-type structure, and one of the antiferromagnets HoGa$_2$ forms the same structure. According to a database\,[19] there are two other compounds in the same form: HoSi$_{1.7}$ and HoGe$_{1.5}$\,[20, 21], but their magnetic properties are not known. Figure\,\ref{fig1} shows the lattice parameter, the magnetic order, and the magnetic entropy changes ($-\Delta S$) reported so far\,[22, 23]. In terms of distance or concentration of Ho$^{3+}$ ions the HoSi$_{1.7}$ and HoGe$_{1.5}$ lie in the intermediate range. They form the layered Ho$^{3+}$ triangular lattice, which raises the question of what magnetic properties they actually have. Likewise their potentially large $-\Delta S$ associated with the geometrical frustration\,[24, 25]. 

We report the synthesis and the magnetic properties of AlB$_2$-type holmium silicides and germanides. Polycrystals were prepared by arc melting, and samples close to the single phase were obtained with the nominal compositions HoSi$_{1.67}$ and HoGe$_{1.67}$. In contrast, melting with the composition HoSi$_{2.00}$ resulted in a dimorph of tetragonal ThSi$_2$-type and orthorombic GdSi$_{1.4}$-type structures. They all showed a decrease of the magnetization below a critical temperature, and the $-\Delta S$ remained similar in amplitude to those of HoGa$_2$\,[23]. Vacancies at the Si or Ge sites can accommodate other elements\,[26] and we attempted to replace vacancies with nickel in order to control the magnetic properties. Nearly single-phase HoGe$_{1.50}$Ni$_{0.50}$ was obtained and its properties were also investigated. \\

\noindent \textbf{Materials and methods} \\
The Ho ingot, Si ingot, Ge powder, and Ni wire were purchased from a commercial vendor and they were reported to be 99.9\,wt\%, 99.999\,wt\%, 99.99\,wt\%, 99\,wt\% pure, respectively. Surface of the Ho ingots were sanded out and the oxide layer was removed, then they were cut into pieces. The pieces were roughly 100-300\,mg in weight, and the amount of other materials was calculated according to the weight. The Si ingot was smashed and the Ni wire was cut. The Ge powder was formed into pellets after weighting. The above processes were carried out in the air. The samples were prepared by arc melting under an Ar atmosphere. The mixtures were turned over and melted several times to ensure homogeneity. 

According to Pukas \textit{et al.}\,[26] the AlB$_2$-type silicides and germanides are stabilized with defects on the Si or Ge sites. A systematic investigation was required to obtain the target crystal structure, and we prepared samples with different compositions HoSi$_{2.00}$, HoSi$_{1.67}$, HoSi$_{1.50}$, HoGe$_{2.00}$, HoGe$_{1.67}$, and HoGe$_{1.50}$. The compositions are used as labels in this report. Powder x-ray diffraction was performed with a commercial diffractometer (Rigaku, MiniFlex600) with Cu-K$\alpha$ radiation ($\lambda$=1.5406\,\AA) and the obtained patterns were analyzed using Rietan package\,[28]. We focused on the identification of crystal phases and lattice parameters. The lattice parameters were stable within the 3 significant figures and they are used to estimate the volume of the samples from the weight. Both the preparation of powder specimens and the measurements were carried out in the air at room temperature. We in addition prepared several pieces with the same recipe and one of the pieces was used for magnetization measurement without shaping. Its shape was nearly spherical, and thus, the demagnetization factor was assumed to be 1/3. All magnetization data reported in this paper are after the demagnetization correction. The dc magnetization was measured with a SQUID magnetometer (Quantum Design, MPMS-XL). We used a straw provided from the company and ensured that background signals were negligible. \\

\begin{figure}[ht]
\centering
\includegraphics[width=\textwidth]{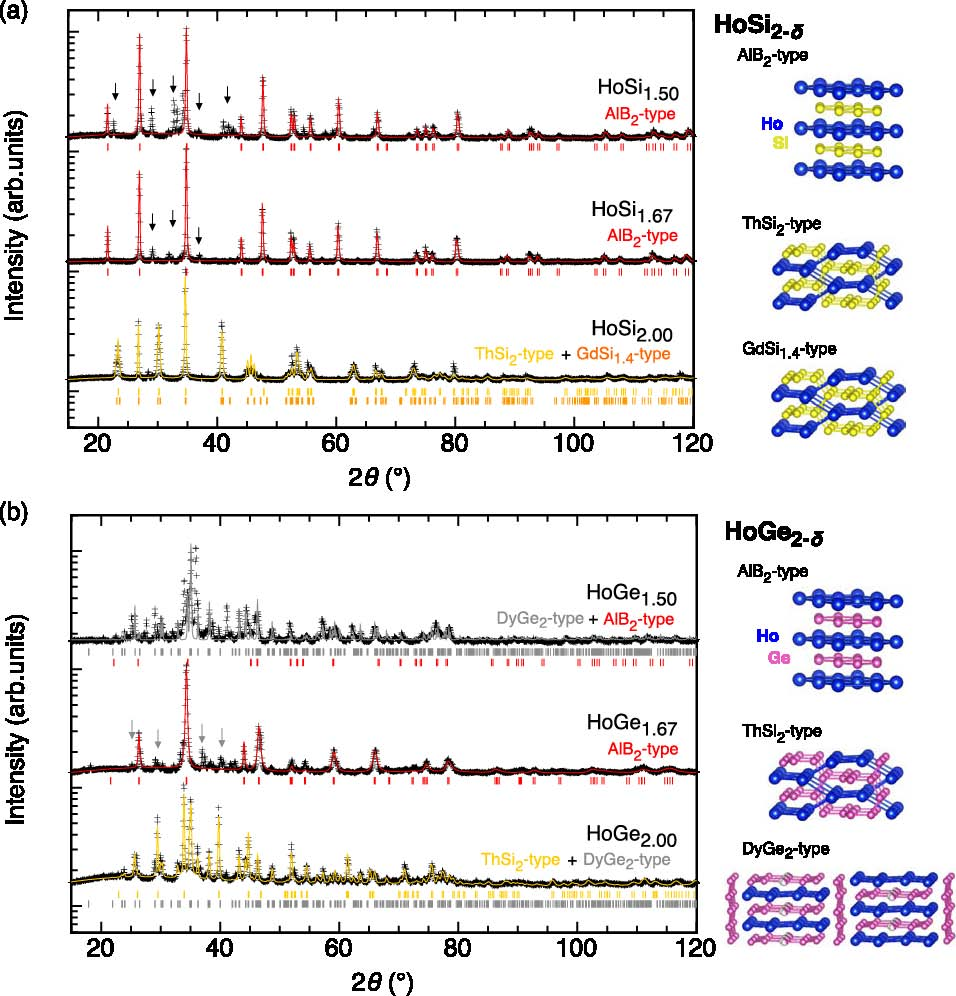}
\caption{\textbf{Powder x-ray diffraction patterns and their fitting.} (a)\,HoSi$_{2-\delta}$ and (b)\,HoGe$_{2-\delta}$. Corresponding crystal images are shown in the right panel. Additional peaks pointed by arrows are from unknown phases.}\label{fig2}
\end{figure}

\noindent \textbf{Results} \\
Figure\,\ref{fig2} shows the powder x-ray diffraction patterns and their fitting. Corresponding crystal images are also shown in the figure\,[27]. Various crystal phases have been reported in the holmium silicides and germanides\,[19], and the HoSi$_{2.00}$ was found to be a dimorph of tetragonal ThSi$_2$-type structure and orthorhombic GdSi$_{1.4}$-type structure. Main phase of the HoSi$_{1.50}$, HoSi$_{1.67}$, and HoGe$_{1.67}$ was found to be the AlB$_2$-type structure. Additional peaks in the HoSi$_{1.50}$ and the HoSi$_{1.67}$ derive from foreign phases, which however could not be identified. Additional peaks in the germanide samples derive from the orthorhombic DyGe$_2$-type structure. The main phases and their lattice parameters are summarized in the Table\,\ref{tab1}.

\begin{figure}[ht]
\centering
\includegraphics[width=\textwidth]{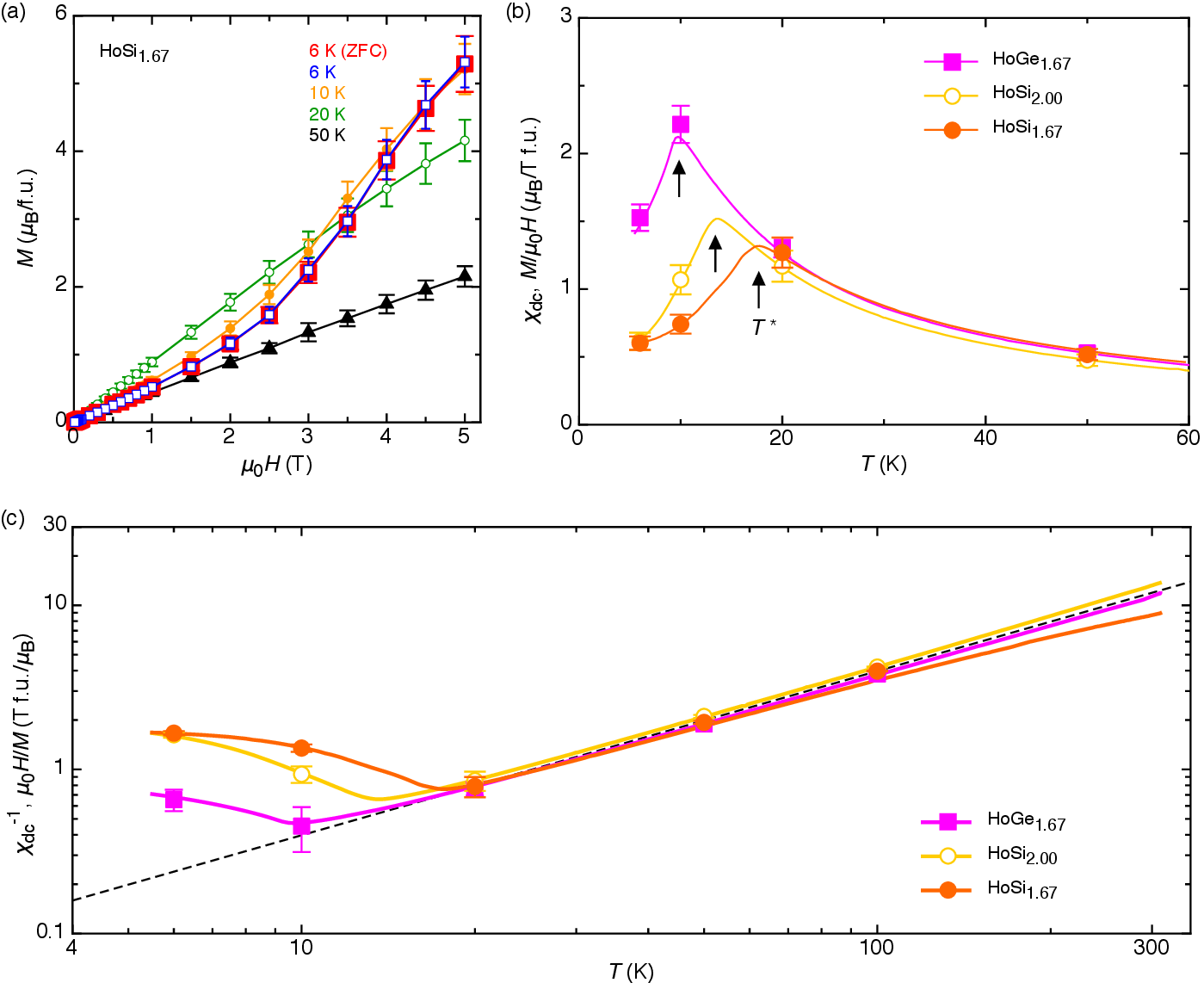}
\caption{\textbf{Magnetization and magnetic susceptibility.} (a)\,$M-H$ curve of the HoSi$_{1.67}$ at selected temperatures. (b)\,Temperature dependence of the magnetic susceptibility. Arrows indicate the critical temperature $T^*$. (c)\,Temperature dependence of the reciprocal susceptibility. The broken line is a simulated curve of the Curie-Weiss model (Eqn.\,\ref{eqn1}) with $\mu_{\rm{eff}}=10.5\mu_{\rm{B}}$, $T_{\rm{CW}}=0$, and $\chi_0=0$.}\label{fig3}
\end{figure}

Magnetic properties of the HoSi$_{1.67}$, HoGe$_{1.67}$, and HoSi$_{2.00}$ were investigated in detail. Figure\,\ref{fig3}(a) shows the magnetic field ($\mu_0H$) dependence of the magnetization ($M$) for the HoSi$_{1.67}$ at selected temperatures. The data at 6\,K were collected with increasing the field after zero-field-cooling (ZFC), and afterwards with decreasing the field.  Magnetic hysteresis was not detected as can be seen in the figure. Nonlinear field dependence was detected below 20\,K. The $M$ of the HoGe$_{1.67}$ and the HoSi$_{2.00}$ were collected using the same measurement sequence, and results were essentially similar (not shown). Figure\,\ref{fig3}(b) shows the temperature ($T$) dependence of the magnetic susceptibility ($\chi_{\rm{dc}}$) at the weak field limit, obtained from the linear fitting of the $M-H$ curve below 1\,T (Fig.\,\ref{fig3}(a)), and the $M/\mu_0H$ obtained from the $M-T$ curve under the field 10\,mT. All HoSi$_{1.67}$, HoGe$_{1.67}$ and HoSi$_{2.00}$ show a steep decrease below a critical temperature $T^*$ at where $\partial (M/H) / \partial T =0$, indicating a suppression of net magnetic moment. We in addition investigated the $T$ dependence with ZFC and FC and confirmed the absence of hystereses in the whole measured temperature range (not shown). These results are consistent with the antiferromagnetic order at the $T^*$ with second-order phase transition.

Figure\,\ref{fig3}(c) shows the same data in the reciprocal susceptibility.  The $\mu_0H/M$ above 95\,K were well approximated by the linear temperature dependence and thus were fit with conventional Curie-Weiss model 
\begin{equation}
\chi_{\rm{dc}} = \frac{\mu_{\rm{eff}}^2}{3 k_{\rm{B}}(T-T_{\rm{CW}})} + \chi_0,
  \label{eqn1}
\end{equation}
where $\mu_{\rm{eff}}$ is the effective magnetic moment, $k_{\rm{B}}$ is the Boltzmann constant, $T_{\rm{CW}}$ is the Curie-Weiss temperature, and $\chi_0$ is the temperature-independent offset which is commonly referred to as the Pauli susceptibility.
There are three open parameters $\mu_{\rm{eff}}$, $T_{\rm{CW}}$, and $\chi_0$, and values obtained from the fitting are listed in the Table\,\ref{tab1}. The $\mu_{\rm{eff}}$ are fairly close to the 10.5$\mu_{\rm{B}}$, but small deviations are found. The HoSi$_{1.67}$ have the largest $\chi_0$ and $T_{\rm{CW}}$ values among the materials investigated in this study. It also showed the highest $T^*$. These results imply an enhancement of magnetic interaction in the HoSi$_{1.67}$.

\begin{figure}[ht]
\centering
\includegraphics[width=\textwidth]{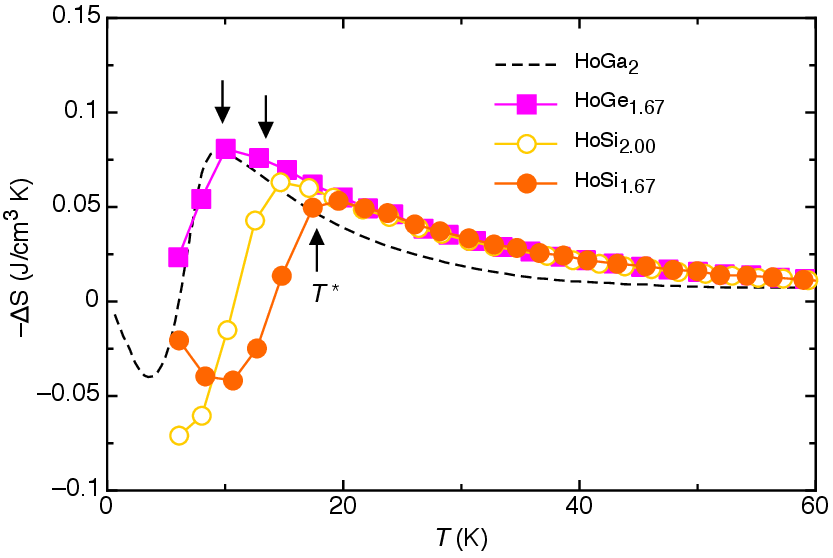}
\caption{\textbf{Magnetic entropy changes from 5\,T to 0\,T.} Arrows indicate the $T^*$. The broken line is the same curve shown in the Fig.\,\ref{fig1}(b).}\label{fig4}
\end{figure}

Figure\,\ref{fig4} shows the $-\Delta S$ from 5T to 0T. They were calculated using the relation 
\begin{equation}
 - \Delta S = -\mu_0 \int^H_0 \Big( \frac{\partial M}{\partial T} \Big)_H dH
  \label{eqn2}
\end{equation}
and the $M-T$ curves in the field 0.1\,T, 0.5\,T, 1\,T, 2\,T, 3\,T, 4\,T, and 5\,T. The $-\Delta S$ of HoGa$_2$ are adopted from\,[23]. Their amplitudes remain moderate due to the change of sign in the $\partial M / \partial T$ at the $T^*$, which are typical of antiferromagnetic phase transition. The $-\Delta S$ at the $T^*$ are listed in the Table\,\ref{tab1}.

We note that the results for the HoSi$_{2.00}$ are overall similar to the previous studies on the HoSi$_{2}$ and the HoSi$_{1.9}$\,[12, 29]. Small deviations in the obtained values are likely to be minor effects on the $-\Delta S$. \\

\begin{table}[ht]
\centering
\scalebox{0.6}{
 \begin{tabular}{rrllllllllll}
   \hline
   Sample & Main phase & $a$ & $c$  & Volume & $\mu_{\rm{eff}}$ & $\chi_0$ & $T_{\rm{CW}}$ & $T^*$ & $- \Delta S$ (5-0\,T) &Reference& \\
    & & (\AA) & (\AA) & (\AA$^3$/f.u.) & ($\mu_{\rm{B}}$) & ($10^{-2}\mu_{\rm{B}}$/T\,f.u.) & (K) & (K) & (J/cm$^3$K) & & \\
   \hline \hline
   HoB$_{2}$ & AlB$_2$-type & 3.2849 & 3.8183 & 35.68 &  - & - & 25 & - & 0.35 &[3, 22] & \\
   HoSi$_{1.50}$ & AlB$_2$-type & 3.81(1) & 4.11(1) & 51.7 & - & - & - & - & - & This work & \\
   HoSi$_{1.67}$ & AlB$_2$-type & 3.82(1) & 4.10(1) & 51.9 & 10.30(4) & 3.37(5) & 6.5(8) & 17.6(2) & 0.05(1) & This work & \\
   HoSi$_{2.00}$ & ThSi$_2$-type/ & 3.97(1) & 13.31(1) & 52.6 & 10.36(1) & -0.47(1) & 1.0(1) & 13.7(2) & 0.06(1) & This work & \\
   HoGe$_{1.67}$ & AlB$_2$-type & 3.91(1) & 4.12(1) & 54.4 & 10.80(1) & 0.17(1) & 1.0(2) & 9.9(2) & 0.08(1) & This work & \\
   HoGe$_{2.00}$ & ThSi$_2$-type & 4.05(1) & 13.67(1) & 56.1 & - & - & - & - & - & This work & \\
   HoGa$_{2}$ & AlB$_2$-type & 4.20 & 4.04 & 61.74 & 10.10(4) & - & -13.1(1) & 8.0(1) & 0.06 & [15, 23] & \\
   \hline
  \end{tabular}
}
\caption{\textbf{Properties obtained from this study.} The $a$, $c$, and volume are from the powder x-ray diffraction, and the $\mu_{\rm{eff}}$, $\chi_0$, $T_{\rm{CW}}$, $T^*$, and $- \Delta S$ are from the dc magnetization. The values for HoB$_2$ and HoGa$_2$ were taken from the literature for comparison.}\label{tab1}
\end{table}

\begin{figure}[ht]
\centering
\includegraphics[width=\textwidth]{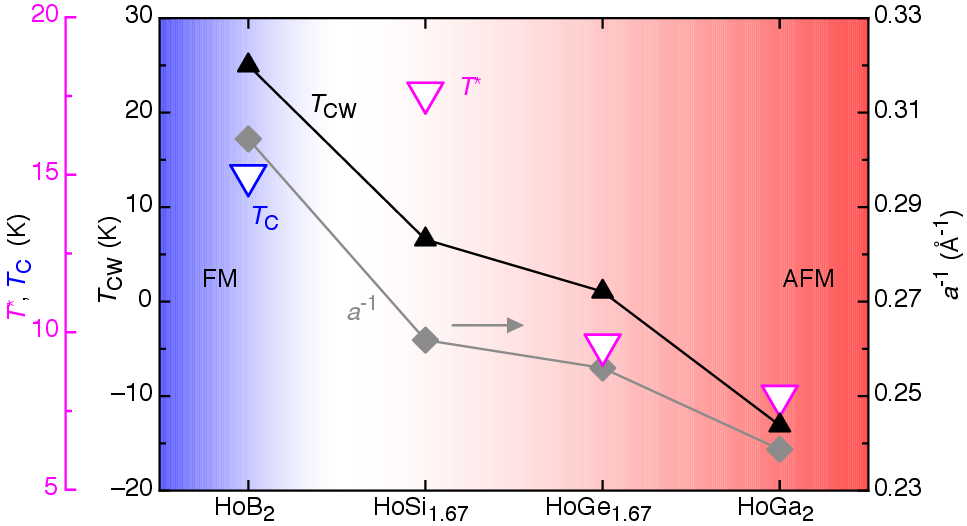}
\caption{\textbf{Magnetic transition temperature and Curie-Weiss temperature.} FM stands for ferromagnetic order and AFM for antiferromagnetic order. Reciprocal of the lattice parameter $a$ is also shown.}\label{fig5}
\end{figure}

\noindent \textbf{Discussion} \\
Various crystal phases have been reported in the rare-earth silicides and germanides, and they involve natural defects at the Si or Ge sites\,[26]. We focused on the AlB$_2$-type structure and the large local moment on the Ho$^{3+}$ ion, and investigated their magnetic properties. They are found to show the antiferromagnetic order, and thus, the present study has revealed that only HoB$_2$ shows the ferromagnetic order in the AlB$_2$-type binary holmium compounds. The $T^*$ which corresponds to the N\'eel temperature, the $T_{\rm{CW}}$, and the Curie temperature ($T_{\rm{C}}$) are shown in the Fig.\,\ref{fig5}. $T_{\rm{CW}}$ represents the strength of the molecular field at high temperature and positive (negative) values are interpreted as ferromagnetic (antiferromagnetic) exchange interaction between the local moments. Reciprocal of the lattice parameter $a$ is in addition shown in the figure; The $a$ corresponds to the minimum distance between the Ho$^{3+}$ ions in the triangular lattice. Interestingly, the $T_{\rm{CW}}$ and the $a$ have similar trends. This implies that the distance between the in-plane Ho$^{3+}$ ions is an important parameter for the magnetic exchange interaction. The $T_{\rm{CW}}$ were positive for both HoSi$_{1.67}$ and  HoGe$_{1.67}$, but their actual orders were antiferromagnetic. The difference indicates a temperature variation of dominant magnetic interaction.

Previous studies have pointed the prevalence of Ruderman-Kittel-Kasuya-Yoshida (RKKY) interaction in both HoB$_2$ and HoGa$_2$\,[15, 22]. It implies the presence of complex magnetic interaction between the local and the itinerant moments, and the same applies to the HoSi$_{1.67}$ and the HoGe$_{1.67}$. Further decrease of the Ho$^{3+}$ distance is most likely favorable for strengthening the ferromagnetic interaction, and thus, the decrease is also likely advantageous to increase the $\partial M/ \partial T$ (Eqn.\,\ref{eqn2}) and the $-\Delta S$. We point out that the RKKY interaction is often considered to be prevalent in other magnetocaloric materials\,[30, 31]. As the distance between the ions are readily available from crystal structure analysis, detailed investigation of them may be useful in the search for other materials.

\begin{figure}[ht]
\centering
\includegraphics[width=\textwidth]{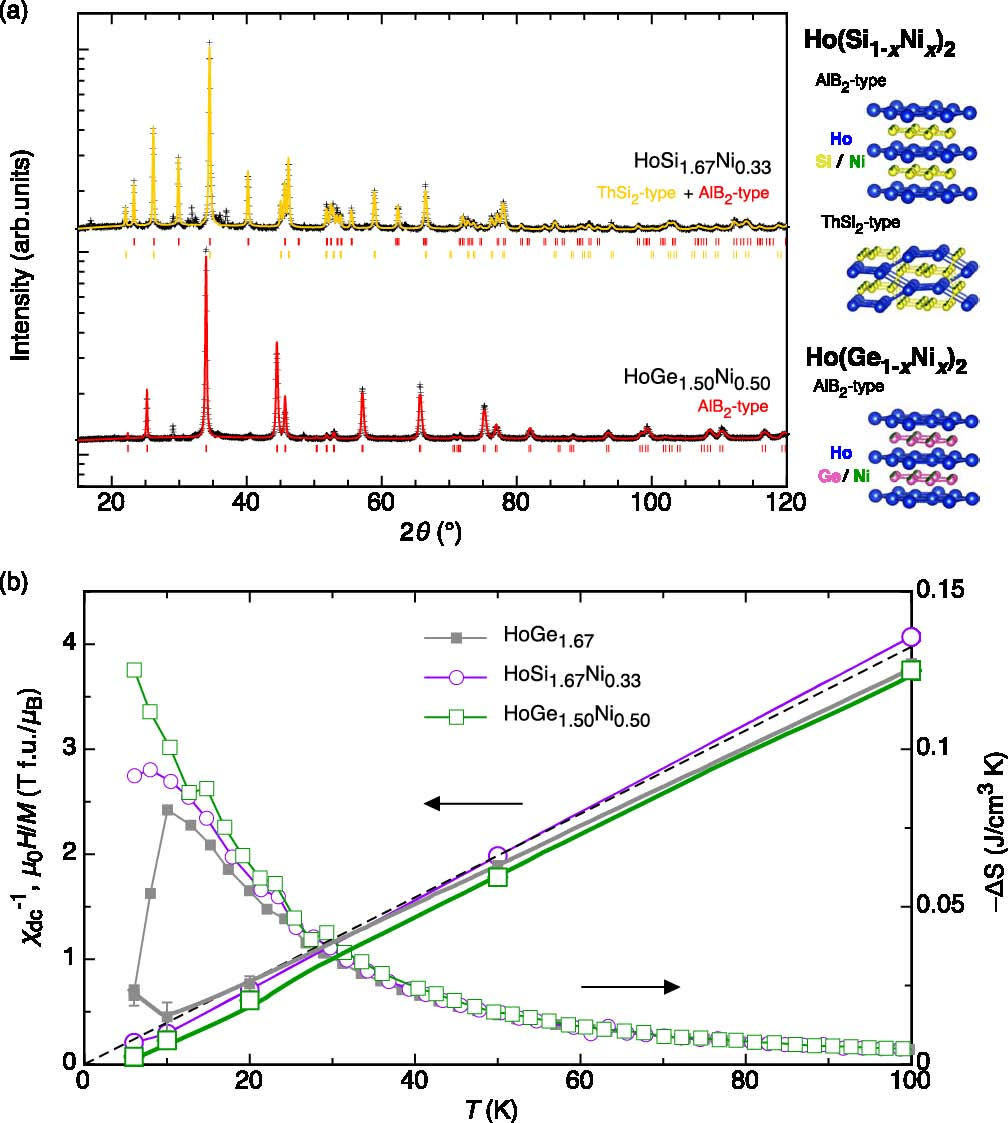}
\caption{\textbf{Substitution of nickel for vacancies at the Si or Ge sites.} (a)\,Powder x-ray diffraction patterns and their fitting for HoSi$_{1.67}$Ni$_{0.33}$ and HoGe$_{1.50}$Ni$_{0.50}$. Corresponding crystal images are shown in the right panel. (b)\,Temperature dependence of the reciprocal susceptibility and the magnetic entropy changes from 5\,T to 0\,T. The data of the HoGe$_{1.67}$ are the same as those shown the Fig.\,\ref{fig3}(c) and Fig.\,\ref{fig4}. The broken line is the same curve shown in the Fig.\,\ref{fig3}(c).}\label{fig6}
\end{figure}

RKKY interaction is the exchange interaction between the local moments mediated by the itinerant spins, and accordingly, it is expected to be sensitive to the spin degree of freedom of the conducting carriers. We considered controlling magnetic properties in this direction. Our results of the sample synthesis appear to support the criterion suggested by Pukas \textit{et al.}\,[26], and the vacancies at the Si or Ge sites can accommodate other elements. The valence electron in the Si or Ge is expected to be conducting, therefore, we have attempted to add polarized itinerant spins by replacing the vacancies with nickel\,[32, 33]. Figure\,\ref{fig6} show the results; A dimorph of ThSi$_2$-type and AlB$_2$-type structures in the nominal composition HoSi$_{1.67}$Ni$_{0.33}$, and a nearly single phase of AlB$_2$-type structure in the HoGe$_{1.50}$Ni$_{0.50}$, are obtained. They did not show magnetic orders down to 5.5\,K and there were increases in the $-\Delta S$ compared to those of HoGe$_{1.67}$. These results support the idea of controlling the local moments using the vacancies, and also underpin the prevalence of RKKY interaction. Properties obtained from this study are listed in the Table\,\ref{tab2}. 

\begin{table}[ht]
\centering
\scalebox{0.7}{
 \begin{tabular}{rrllllllll}
   \hline
   Sample & Main phase & $a$ & $c$  & Volume & $\mu_{\rm{eff}}$ & $\chi_0$ & $T_{\rm{CW}}$ &Reference& \\
    & & (\AA) & (\AA) & (\AA$^3$/f.u.) & ($\mu_{\rm{B}}$) & ($10^{-2}\mu_{\rm{B}}$/T\,f.u.) & (K) & & \\
   \hline \hline
   HoSi$_{1.67}$Ni$_{0.33}$ & AlB$_2$-type & 3.93(1) & 4.02(1) & 53.8 & - & - & -  & This work & \\
   HoSi$_{1.67}$Ni$_{0.33}$ & ThSi$_2$-type & 3.98(1) & 13.59(1) & 53.7 & - & - & - & This work & \\
   HoGe$_{1.50}$Ni$_{0.50}$ & AlB$_2$-type & 4.08(1) & 3.97(1) & 57.1 & 10.74(1) & -0.12(1) & 4.7(2) & This work & \\
   \hline
  \end{tabular}
}
\caption{\textbf{Properties obtained from this study.} The $a$, $c$, and Volume are from the powder x-ray diffraction, and the $\mu_{\rm{eff}}$, $\chi_0$, and $T_{\rm{CW}}$ are from the dc magnetization.}\label{tab2}
\end{table}

Finally, let us not omit mentioning the magnetic structure in the ordered states. So far, an in-plane antiferromagnetic order in HoGa$_2$ and an out-of-plane ferromagnetic order in HoB$_2$ have been reported\,[15, 22]. Further insights of actual magnetic structures shall be investigated by neutron diffraction. \\

\noindent \textbf{Conclusion} \\
In summary, we investigated magnetic properties of the AlB$_2$-type binary holmium silicides and germanides. They are found to show the antiferomagnetic order and the magnetic entropy changes were similar to those of HoGa$_2$. The magnetic order was suppressed by substituting nickel for vacancies, and there were increases in the magnetic entropy changes. A correspondence is found between the Curie-Weiss temperature and the in-plane Ho$^{3+}$ distance. RKKY interaction is suggested as the prevailing interaction and a small Ho$^{3+}$ distance is likely favorable to strengthen the ferromagnetic interaction and to increase magnetic entropy changes. \\

\noindent \textbf{Funding sources} \\
Authors acknowledge the financial support from the JST-Mirai Program (Grant No. JPMJMI18A3) and JSPS KAKENHI (Grant Nos. 20K05070, 23K04572, 19H02177).

\clearpage
\bibliographystyle{unsrt}

\end{document}